\documentclass[conference]{IEEEtran}
\IEEEoverridecommandlockouts
\usepackage{cite}
\usepackage{amsmath,amssymb,amsfonts}
\usepackage{algorithmic}
\usepackage{graphicx}
\usepackage{textcomp}
\usepackage{multirow}
\usepackage{balance}
\usepackage{hyperref}
\usepackage{xcolor}
\usepackage{booktabs, tabularx}
\usepackage{verbatim}
\def\BibTeX{{\rm B\kern-.05em{\sc i\kern-.025em b}\kern-.08em
    T\kern-.1667em\lower.7ex\hbox{E}\kern-.125emX}}
\begin{document}

\title{Fast Depth Estimation for View Synthesis
\thanks{This work was supported by Innovate UK: Palantir -- Real time inspection and assessment of wind turbine blade health.}
}

\author{\IEEEauthorblockN{Nantheera Anantrasirichai}
\IEEEauthorblockA{\textit{Visual Information Laboratory} \\
\textit{University of Bristol}\\
Bristol, UK \\
n.anantrasirichai@bristol.ac.uk}
\and
\IEEEauthorblockN{Majid Geravand}
\IEEEauthorblockA{\textit{Braendler Engineering}\\
Turin, Italy\\
majid.geravand@braendler.com}
\and
\IEEEauthorblockN{David Braendler}
\IEEEauthorblockA{\textit{Braendler Engineering}\\
London, UK\\
david.braendler@braendler.com}
\and
\IEEEauthorblockN{David R. Bull}
\IEEEauthorblockA{\textit{Visual Information Laboratory} \\
\textit{University of Bristol}\\
Bristol, UK \\
dave.bull@bristol.ac.uk}

}

\maketitle

\begin{abstract}
Disparity/depth estimation from  sequences of stereo images is an important element in 3D vision. Owing to occlusions, imperfect settings and homogeneous luminance,  accurate estimate of depth remains a challenging problem. Targetting view synthesis, we propose a novel learning-based framework making use of dilated convolution, densely connected convolutional modules, compact decoder and  skip connections. The network is shallow but dense, so it is fast and accurate. Two additional contributions -  a non-linear adjustment of the depth resolution and the introduction of a projection loss, lead to reduction of estimation error by up to 20\% and 25\% respectively.
The results show that our network outperforms state-of-the-art methods  with an average  improvement in accuracy  of depth estimation and view synthesis by approximately 45\% and 34\% respectively. Where our method generates comparable quality of estimated depth, it performs 10 times faster than those methods.

\end{abstract}


\section{Introduction}

In the human visual system, a  stereopsis process creates a perception of three-dimensional (3D) depth from the combination of the two spatially separated signals received by the brain from our eyes. The fusion of these two slightly different pictures gives the sensation of strong three-dimensionality  by matching  similarities. To provide stereopsis in machine vision applications, two images are captured simultaneously from two  cameras with parallel camera geometry, and an implicit geometric process is used to extract 3D information from these images. Binocular disparity $d$ is computed and depth $z$ is obtained from (\ref{eqn:z}), 
\begin{equation}
\label{eqn:z}
	 z=\frac{f B}{d}.
\end{equation}
\noindent where $f$ and $B$ are a focal length and a baseline between two cameras, respectively.

3D information, or depth, is utilised in many applications, including 3D reconstruction \cite{Zhou:semi:2018}, view synthesis \cite{Fickel:disparity:2017}, object recognition \cite{Wang:object:2008} and multi-view video compression \cite{Anantrasirichai:inband:2010}. Traditional methods search the corresponding points between the left and the right images using block-based \cite{Tzovaras:Evaluation:1994, Anantrasirichai:inband:2010}  or mesh-based matching \cite{Fickel:disparity:2017}. More sophisticated approaches, e.g. dynamic programming \cite{Wang:object:2008, Anantrasirichai:dynamic:2006}, produce better results as they do not introduce blocking artefacts or noisy depth maps.
Most methods however involve an iterative process to minimise an error function, to further refine the depth map, particularly around the edge of the object \cite{Fan:Road:2018}, and to improve geometric projection \cite{Ince:depth:2007}.  Such iterations are time-consuming and are not suitable for real-time applications. For example, a 3D patch-based minimum spanning tree (3DMIST \cite{Li:3DMIST:2017}), one of the top five in Middlebury Stereo benchmark \cite{Scharstein:taxonomy:2002}, takes about 25~sec to process one 450$\times$350 image pair. On the KITTI benchmark, where the stereo pairs are captured in driving scenes \cite{Menze:Object:2015}, convolutional neural networks (CNNs) have shown significantly faster computation ($<$2.5~sec at the resolution of  1392$\times$512 pixels). However, the estimated disparity maps are mainly used for visual odometry, 3D object detection and 3D tracking, not for 3D reconstruction or view synthesis, where precise estimation at depth discontinuities is crucial.

In this paper, we present a new learning-based approach that achieves both high quality estimated depth for view synthesis and fast computation. We adapt a DenseMapNet \cite{Atienza:Fast:2018} with additional compact decoder and skip layers to include the low-level features for finer estimation. Therefore the network is significantly shallower than many state-of-the-art methods, whilst producing comparably accurate depth results.
Stereo matching is difficult in homogeneous areas and traditional methods solve this issue by using large windows. In the CNN, this can be solved using a big receptive field, so we employ  several dilation rates  to capture disparity. As our depth estimation method is intended to be used for view synthesis,
we herein propose an exponential adjustment for depth values during training process. This will concentrate more on the near objects, which are more salient than the far ones and the background. In addition, we propose a projection loss, where the weights in the convolution layers are also adjusted according to the error from the synthesised image.

The remainder of this paper is organised as follows.  Related work on CNN-based depth estimation is presented in Section \ref{sec:relatedwork}. The proposed scheme  is described in Section \ref{sec:proposed}. The performance of the method is evaluated in Section \ref{sec:results}. Finally, Section \ref{sec:conclusion} presents the conclusions of this work.

\begin{figure*}[t!]
	\centering
 		\includegraphics[width=\textwidth]{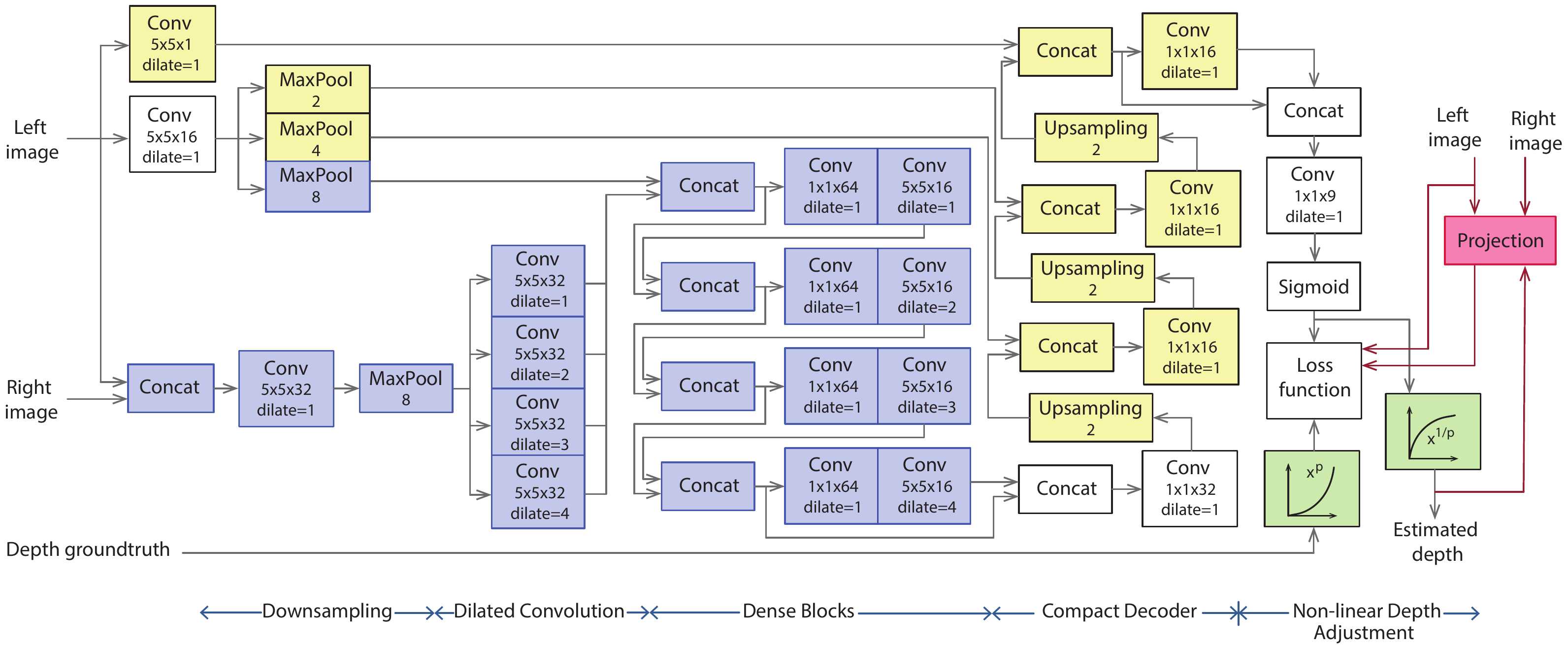}
	\caption{Proposed network architecture for depth estimation. The blue blocks are for feature matching, the yellow blocks are additional compact decoder part to ensure sharp upsampled feature maps, and the green blocks enhance depth resolution of the near objects. Each Conv module comprises convolution, batch normalisation, and ReLu layers.}
	\label{fig:network_structure}
\end{figure*}

\section{CNNs for disparity/depth estimation}
\label{sec:relatedwork}

CNNs were first introduced for stereo matching by Zbontar and LeCun \cite{Zbontar:Stereo:2016} to replace the computation of the matching cost and to learn a similarity measure on small image patches. This method significantly accelerates the process of disparity estimation; as a result, most recent methods employ CNNs. Generally, the learning-based disparity estimation methods comprise two modules: i) \textit{feature extraction}, applied to the left and the right image pair separately, but with the learnable weights and biases shared between them in the training process (e.g. siamese networks), ii) \textit{disparity estimation}, where the output of the feature extraction process, \textit{a cost volume},  is employed to compute a disparity map.
The 3D cost volume stores the costs for choosing a disparity value for each pixel. The simplest cost is the absolute differences of the intensities. The size of the cost volume thus increases with the search space.

As the accuracy of the estimated disparity is significantly improved and  real-time computation becomes feasible, CNNs have been gained more attention. A number of network architectures have been proposed, along with slight changes of parameters and transfer learning for specific applications. For example, a pyramid stereo matching network (PSMNet) \cite{Chang:pyramid:2018} exploits global context information in stereo matching using pyramid pooling module and dilated convolution is applied to further enlarge the receptive field.  It employs a stacked hourglass architecture \cite{Newell:hourglass:2016} to estimate disparity values.
The Sparse Cost Volume Network (SCV-Net)  \cite{Lu:sparse:2018} was proposed to improve complexity efficiency by shifting feature maps with a stride of 3.
Guided Aggregation Net (GA-Net) \cite{Zhang:GANet:2019} employs a stacked hourglass CNN to extract features of the left and right image pair, giving the output as a 4D cost volume. The cost aggregation module  then calculates a disparity map.

% -------------------------------------------------
\section{Proposed method}
\label{sec:proposed}

\subsection{Network architecture}
The proposed network architecture is shown in Fig. \ref{fig:network_structure}.
We adapt the correspondence network from DenseMapNet  \cite{Atienza:Fast:2018} (blue blocks in Fig. \ref{fig:network_structure}) to find correspondences of the input stereo pair. The matching process of corresponding points between the left and the right views is performed at a lower resolution, and down-sampled via a max-pooling layer (a downscale factor of 8 is used throughout this paper). This improves  computation speed, reduces  memory requirements, and overcomes problems of large disparities. Then, dilated convolution with different dilation factors ($l$=1-4) is applied.   Dilated convolution enlarges the field of view of the filters to incorporate larger context by  expanding the receptive field without loss of resolution. The dilated convolutions are defined  in Eq.\ref{eqn:dilatedconv} \cite{Yu:Multi:2016}, where $F$ is a feature map, $k$ is a filter, $\ast_l$ is a convolution operator with a dilation factor $l$. 
\begin{equation}
\label{eqn:dilatedconv}
	 (F\ast_l k)(\textbf{p}) = \sum_{\textbf{s} + l \textbf{t} = \textbf{p}} F(\textbf{s}) k(\textbf{t}).
\end{equation}

Four one-layer Dense Blocks \cite{Huang:Densely:2017} are then employed to capture corresponding features. A Dense Block uses  feature maps from multiple preceding layers as inputs leading to more connections amongst layers. Subsequently, the feature maps are enlarged to the original resolution. Instead of applying upsampling only once like in DenseMapNet, we propose a compact decoder using a step-wise upsampling of 2 (yellow blocks in Fig. \ref{fig:network_structure}). In addition, we add  skip connections by merging the low-level feature maps of the left image in every upsampling step. This ensures pixel-wise co-locations between the RGB image and the depth map in both full resolution and feature levels.

If the ground truth data for training is based on disparity values, it is scaled to $[0,1]$, equivalent to $[0, d_{\max}]$, where $d_{\max}$ is the maximum value of disparity. If the network is to estimate depth, the ground truth $\hat{z}$ is normalised and subtracted from 1, i.e. $\hat{z}= 1-\frac{z}{z_{\max}}$. The prediction output of the network is done via a Sigmoid activation function, $\hat{x}=(1 + e^{-x})^{-1}$, which scales the output $x$ of the last convolution layer to $[0,1]$.

\subsection{Depth adjustment}
For view synthesis, foreground objects are often more salient and incorrect depth estimates can result in noisy visualisation, particularly at the edges of the objects or where  there exists discontinuity of the depth. Here we propose a nonlinear adjustment to the depth ground truth (green blocks in Fig. \ref{fig:network_structure}) so that the closer objects have higher depth resolution than distant objects or background. During the training process, exponentiation is applied to the ground truth $\hat{z}$, as  in (\ref{eqn:expo}), where $p$ is an exponent. For the prediction process, the final estimated depth $ \tilde{z}$ is computed using  (\ref{eqn:reexpo}), where $\hat{x}$ is the output of the Sigmoid activation layer.
\begin{equation}
\label{eqn:expo}
	 \hat{z}' = \hat{z}^p, \: \hat{z} \in [0,1] \: \text{and} \: p \geqslant 1.
\end{equation}
\begin{equation}
\label{eqn:reexpo}
	 \tilde{z} = \hat{x}^{\frac{1}{p}}, \: \hat{x} \in [0,1].
\end{equation}

\begin{figure}[htbp]
	\centering
 		\includegraphics[width=\columnwidth]{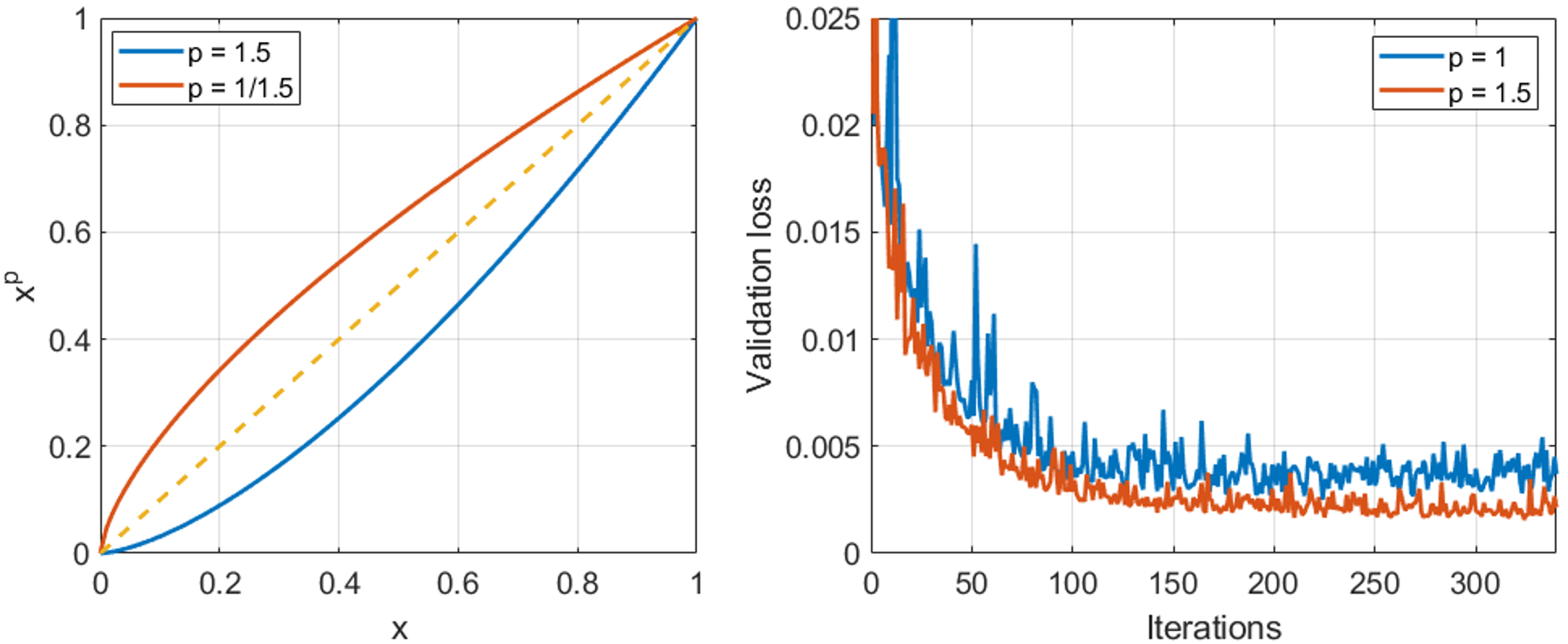}
	\caption{Exponential functions (left) and validation losses with and without depth adjustment.}
	\label{fig:exponentiation} 
\end{figure}

Fig. \ref{fig:exponentiation} (left) demonstrates how $\hat{z}$ values are adjusted with the exponential function when $p$=1.5. The blue plot shows that $\hat{z}$ values close to 1 (areas near to the cameras) are stretched out gaining higher resolution, whilst the values close to 0 are shrunk (areas far  from the cameras). This technique improves the validation loss  by approximately 15\%, as shown in Fig. \ref{fig:exponentiation} (right). The optimal value of $p$ depends on applications and the positions of the salient objects in the scene. We initialise the $p$ value using  curve fitting to the histogram of the training ground truth.

\subsection{Loss function}
Most networks for disparity estimation mentioned in Section \ref{sec:relatedwork} employ a smooth $\ell_1$ loss function (mean absolute error) as it is robust to outliers and disparity/depth discontinuities. However, outliers can still produce undesirable projected pixels - seen as noisy edges in the synthesised views. Therefore we employ the $\ell_2$ loss function (mean square error). Note that we tested several loss functions, including cross entropy, $\ell_1$, $\ell_2$ and perceptual loss with pre-trained VGG \cite{Johnson:Perceptual:2016}. While the qualities of estimated depths do not differ significantly, the best view synthesis is accomplished with the $\ell_2$ loss.

In this paper, we propose using the $\ell_2$ losses calculated from the predicted depth map $L_{\ell_2}$ and the reconstructed left image from the right image, referred as the projected loss $L_{R\rightarrow L}$. The prediction loss $L_{\ell_2}$ equals to $\sum (\hat{z} - z)^2$, where $z$ and $\hat{z}$ are real and estimated depth values. For the projection loss, if needed, the depth is first converted to  disparity $d$ through the relationship in (\ref{eqn:z}), i.e. $\hat{d} = {f B}/{\hat{z}}$. The pixel ($i$, $j$) on the reconstructed left image is derived from the pixel ($i-\hat{d}$, $j$) on the right image. The final loss function is a weighted combination between two losses as in (\ref{eqn:loss}), where $\alpha_z$ and $\alpha_p$ are the weights of the prediction and projection losses (We simply use $\alpha_z$=$\alpha_p$=1 in this paper).  $I^L$,  $I^R$, ${N_z}$,  and ${N_p}$ are the left image, the right image, the total number of the pixels on each image, and the total number of existing pixels on the reconstructed left image, respectively. Experimental results show that adding the projected loss improves the prediction performance by approximately 20\%, compared to using $L_{\ell_2}$ alone.
\begin{align}
\label{eqn:loss}
\begin{split}
	 L &= \alpha_z L_{\ell_2} + \alpha_p L_{R\rightarrow L} \\
	  &= \frac{1}{N_z} \sum_k (\hat{z}_k - z_k)^2 + \frac{1}{N_p} \sum_{ij} (I^R_{i-\hat{d},j} - I^L_{ij})^2
\end{split}
\end{align}

% -------------------------------------------------
\section{Results and discussion}
\label{sec:results}

\begin{figure}[htbp]
	\centering
 		\includegraphics[width=\columnwidth]{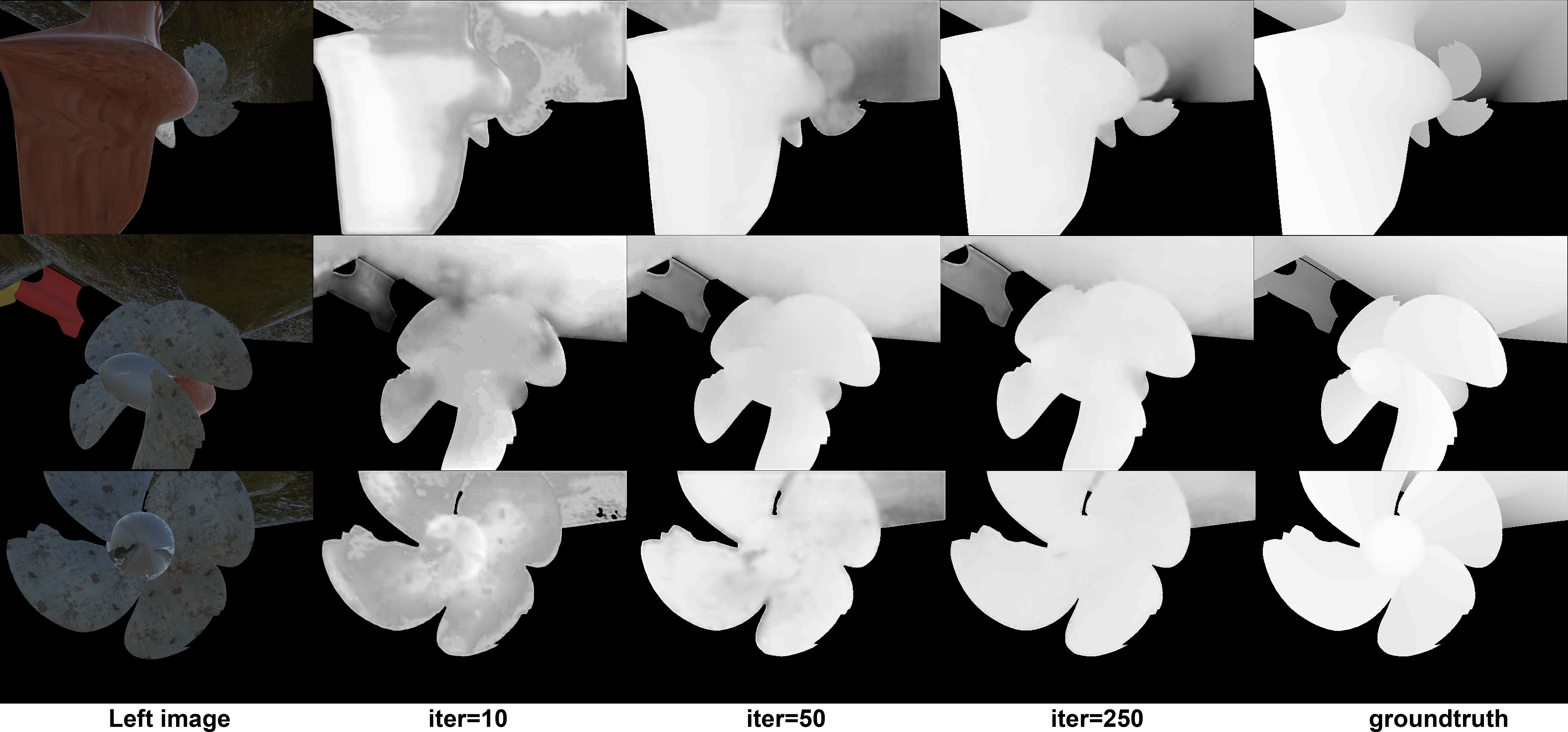}
	\caption{Estimated depth maps of \textit{Propeller} dataset at the iterations of 10, 50 and 250 (column 2--4) of the stereo pair, where the left image and the true depth are shown in the first and the last column, respectively.  Note that the intensity is normalised and adjusted for visualisation.}
	\label{fig:results_iter}
\end{figure}

The proposed network is implemented on Keras with a Tensorflow backend (available at \url{https://github.com/pui-nantheera/DepthEstimation}). The network is trained with the Adam optimizer, which is an extension to stochastic gradient descent (SGD), in which the procedure updates network weights iteratively based on training data. To prevent overfitting, we drop 20\% of features extracted in the convolution layers.
The model is initialised by training the network with our synthetic datasets, described in Section \ref{ssec:syndata}, and subsequently using it as the initial model when training with other test stereo sequences. This transfer learning strategy by fine-tuning a pre-trained network speeds up the training process. Referring the validation loss of $p$=1.5 in Fig. \ref{fig:exponentiation}, the estimated depth maps during the training are shown in Fig. \ref{fig:results_iter}. The depth is quickly learnt, getting sharper and settles around the 250$^{\text{th}}$ iteration. Experimental results and state-of-the-art comparison of depth estimation and view synthesis are presented in Section \ref{ssec:resultcompare}.

\subsection{Synthetic datasets}
\label{ssec:syndata}

\begin{figure}[htbp]
	\centering
 		\includegraphics[width=\columnwidth]{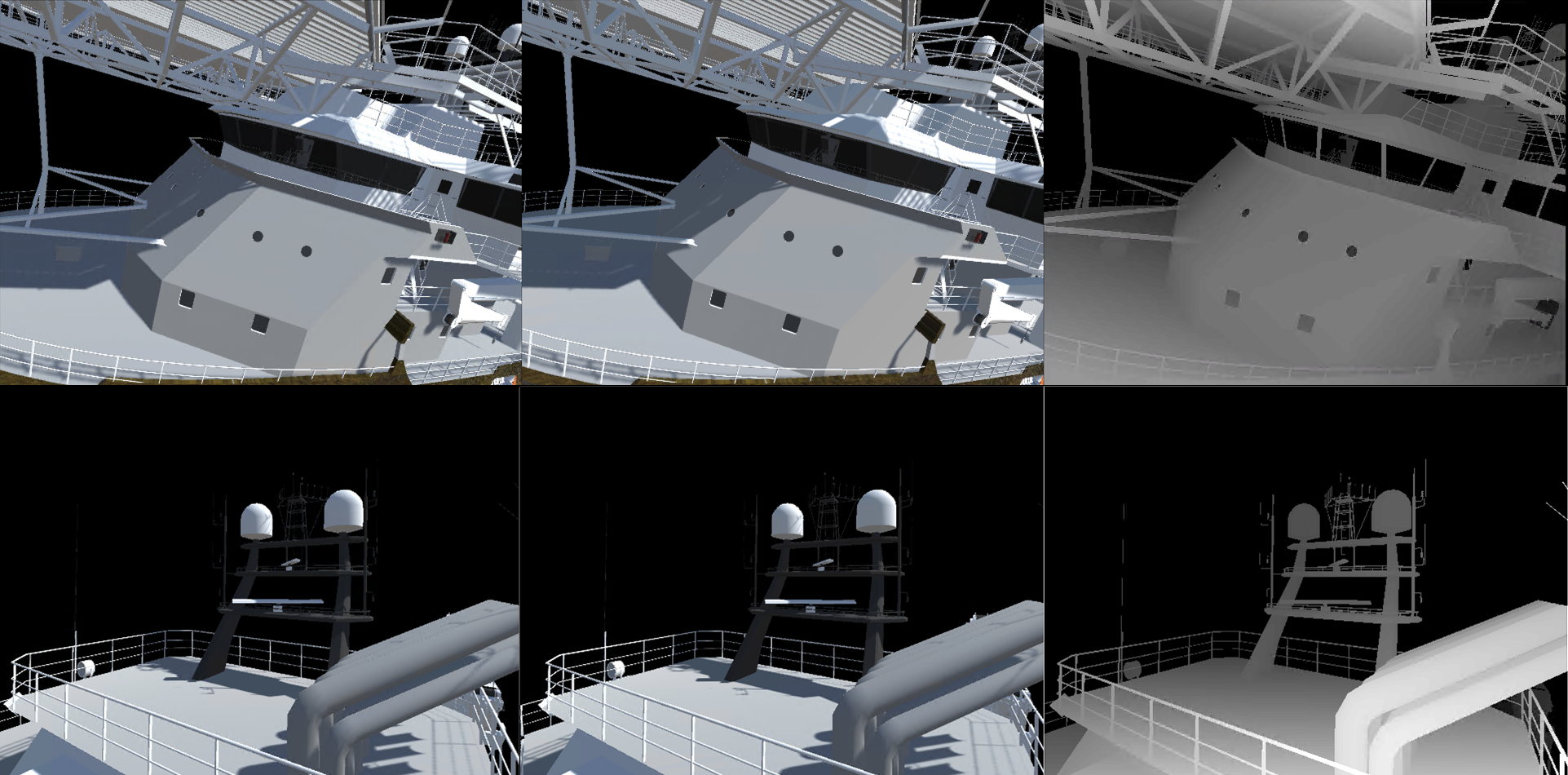}
	\caption{Stereo pairs of `\textit{Ship}' and `\textit{Antenna}' scenes in the top and bottom rows, respectively. The depth maps are shown in the last column where the intensity represents the depth -- brighter indicating closer to the camera.}
	\label{fig:ship_stereo}
\end{figure}

For robustness, we include  eight scenes of simulated stereo sequences (total 25,000 stereo pairs with a resolution of 480$\times$640 pixels), created using Unity software \cite{Unity:2019} and the Elastic Fusion algorithm \cite{Whelan:ElasticFusion:2016} by Braendler Engineering\footnote{http://braendler.com/home}. Three cameras are attached together. Two of them are identical and used to capture RGB images with parallel camera configurations, whilst the other captures depth images (range 0-15m).  The depth images are recorded, corresponding to the left image. All control parameters are set to replicate the real scenario. The examples of the synthetic datasets are illustrated in Fig.  \ref{fig:results_iter} and \ref{fig:ship_stereo}. The scenes include both simple and complicated structures, with narrow objects at different depths. The cameras are moved around the target generating various values. These datasets are available on \url{https://go.aws/37zlsTs}.

\subsection{Performance}
\label{ssec:resultcompare}

We tested our network with standard test sequences, namely 
i) \textit{Sintel} \cite{Butler:Naturalistic:2012}, containing animated humans, animals in various types of background scenes, ii) \textit{Driving} and \textit{Monkaa} \cite{Mayer:large:2016}, containing different sizes of objects and visually challenging fur. The \textit{Driving} scenes replicate the KITTI2015 dataset \cite{Menze:Object:2015}, but provide dense disparity groundtruth. We also include one of our synthetic datasets, \textit{Propeller}, for testing. For this dataset, the depth maps are estimated, instead of disparity maps. We randomly selected 90\% of the stereo pairs for training and used the remainder for testing. 

The performance of the proposed network was compared with three state-of-the-art methods, i.e. DispNet \cite{Mayer:large:2016}, DenseMapNet \cite{Atienza:Fast:2018}, and PSMNet \cite{Chang:pyramid:2018}. The results of the depth/disparity estimation are shown in Table \ref{tab:result}.
We also compared the quality of synthesised views generated using the estimated depth or disparity. Table \ref{tab:result} shows the mean absolute error (MAE) of the synthesised right images projected from the left images. The results show that our network achieves the best performance on the \textit{Driving} and \textit{Propeller} datasets, whilst the PSMNet outperforms others on the \textit{Sintel} and \textit{Monkaa} datasets. However, when comparing the results of view synthesis, our network and PSMNet shows an insignificant difference on the \textit{Sintel} and \textit{Monkaa} datasets, but our network outperforms the PSMNet by approximately 20\% and 10\% on he \textit{Driving} and \textit{Propeller} datasets, respectively. This is because our method places greater emphasises on the foreground.

\begin{table}[htbp]
\caption{Performance comparison showing mean end-point-error (EPE) of depth/disparity estimation ($\hat{z}$) and mean absolutue error (MAE) of reconstructed right view  ($\hat{I}^R$). The EPEs of Sintel, Driving, and Monkaa are in pixels, whilst those of Propeller are in cm.}
\begin{center}
\scriptsize
\begin{tabular}{|c|c|c|c|c|c|c|c|c|}
\hline
\multirow{2}{*}{\textbf{Method}}&\multicolumn{2}{|c|}{\textbf{\textit{Sintel}}}& \multicolumn{2}{|c|}{\textbf{\textit{Driving}}}& \multicolumn{2}{|c|}{\textbf{\textit{Monkaa}}} & \multicolumn{2}{|c|}{\textbf{\textit{Propeller}} } \\
\cline{2-9} 
 & $\hat{z}$& $\hat{I^R}$& $\hat{z}$& $\hat{I^R}$& $\hat{z}$& $\hat{I^R}$& $\hat{z}$& $\hat{I^R}$\\
\hline
DispNet & 5.38 & 9.87 & 15.62  & 13.93 & 5.99 & 12.84 & 9.73  & 59.34   \\
\hline
DenseMapNet & 4.41 &  8.34 &  6.56 & 9.43  & 4.45 & 9.78 & 8.67  &  54.69 \\
\hline
PSMNet          & \textbf{3.85 } &  \textbf{7.94} & 8.12   &  10.32 & \textbf{3.88} & \textbf{7.36} & 2.38 & 41.22   \\
\hline
Proposed       & 3.95  & 7.99   &  \textbf{6.42}  & \textbf{8.24 } & 4.08 &  7.93 & \textbf{2.32} & \textbf{37.31}  \\
\hline
\end{tabular}
\label{tab:result}
\end{center}
\end{table}

Amongst the state-of-the-art methods, DenseMapNet was reportedas having the fastest speed (less than 0.03 sec per stereo pair using NVIDIA GTX 1080Ti) \cite{Atienza:Fast:2018,Mayer:large:2016,Chang:pyramid:2018}. This is followed by  DispNet and  PSMNet, of which the runtimes are approximately twice and ten times that of DenseMapNet, respectively. We tested DenseMapNet on an NVIDIA Tesla M60 (which is less powerful than NVIDIA GTX 1080Ti) and found that it has a computational time of  0.20 sec per image pair. Our network processes one stereo pair within 0.21 sec which is only 5\% slower than the DenseMapNet, whilst the  quality improvement is in excess of 70\%. Comparing to  PSMNet, the accuracies of disparity estimation for  \textit{Sintel} and  \textit{Monkaa} datasets are 4\% more than that of our method, whilst the performance is about 2\% worse when estimating depths of \textit{Propeller}. This could be because the PSMNet is designed for disparity estimation and the precision of the results is limited to 1 pixel. In contrast, our method can estimate the depth values directly, which means the precision can be up to 64 bits using double-precision floating-point format. Note that  PSMNet takes more than 2 sec per stereo pair using NVIDIA GTX 1080Ti platform, which is 10 times more than our network, whilst realising comparable accuracy.

\begin{figure*}[htbp]
	\centering
	\includegraphics[width=\textwidth]{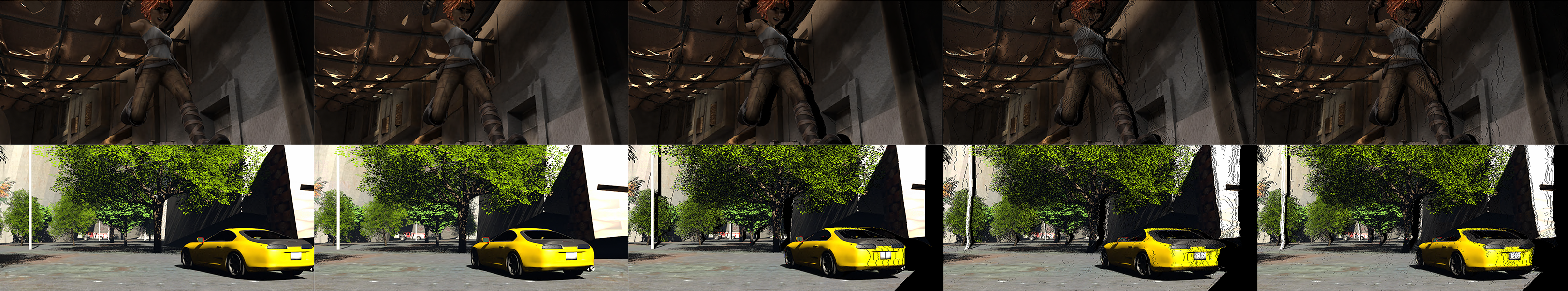}
	\begin{tabular}{p{10mm}p{30mm}p{32mm}p{32mm}p{33mm}p{32mm}} 
		& \small (a) & \small (b) & \small (c) & \small (d) & \small (e) \\
	\end{tabular}
	\caption{View synthesis of \textit{Sintel} and \textit{Driving}. (a) Left image. (b) Right image. (c) Reconstructed right image using groundtruth (d) Reconstructed right image using  estimated depth map by DenseMapNet \cite{Atienza:Fast:2018}. (e) Reconstructed right image using  estimated depth map by our method.}
	\label{fig:reconright}
\end{figure*}

The synthesised right views are shown in Fig. \ref{fig:reconright}. The images are generated from the left images using the groundtruth disparity maps, and those estimated from DenseMapNet \cite{Atienza:Fast:2018} and our method. Our synthesised results show better quality, particularly around the edges of the objects.

% -------------------------------------------------
\section{Conclusions}
\label{sec:conclusion}

This paper presents a new architecture to estimate depth or disparity targetted at view synthesis applications. The proposed network is employs densely connected convolutional modules and dilated convolutions. We improve the estimation by adding a compact decoder and connecting it with the low-level features via skip layers. Additionally, we modify the depth values during  training so that the near objects, which are more salient, have better depth resolution. We also propose a new loss function that integrates a projection loss to maximise the quality of view synthesis. The results show an improvement in prediction accuracy by up to 45\%. Moreover, it significantly improves the perceived quality of 3D reconstruction.

\balance
\bibliographystyle{IEEEbib}
\bibliography{literature_review}

\begin{thebibliography}{10}

\bibitem{Zhou:semi:2018}
Yi~Zhou, Guillermo Gallego, Henri Rebecq, Laurent Kneip, Hongdong Li, and
  Davide Scaramuzza,
\newblock ``Semi-dense 3d reconstruction with a stereo event camera,''
\newblock in {\em European Conf. on Computer Vision (ECCV)}, 2018.

\bibitem{Fickel:disparity:2017}
Guilherme~P. Fickel and Claudio~R. Jung,
\newblock ``Disparity map estimation and view synthesis using temporally
  adaptive triangular meshes,''
\newblock {\em Computers \& Graphics}, vol. 68, pp. 43 -- 52, 2017.

\bibitem{Wang:object:2008}
{Yizhou Wang}, M.~{Brookes}, and P.~L. {Dragotti},
\newblock ``Object recognition using multi-view imaging,''
\newblock in {\em 2008 9th International Conference on Signal Processing}, Oct
  2008, pp. 810--813.

\bibitem{Anantrasirichai:inband:2010}
N.~{Anantrasirichai}, C.~N. {Canagarajah}, D.~W. {Redmill}, and D.~R. {Bull},
\newblock ``In-band disparity compensation for multiview image compression and
  view synthesis,''
\newblock {\em IEEE Transactions on Circuits and Systems for Video Technology},
  vol. 20, no. 4, pp. 473--484, April 2010.

\bibitem{Tzovaras:Evaluation:1994}
Dimitrios Tzovaras, Michael~G. Strintzis, and Haralambos Sahinoglou,
\newblock ``Evaluation of multiresolution block matching techniques for motion
  and disparity estimation,''
\newblock {\em Signal Processing: Image Communication}, vol. 6, no. 1, pp. 59
  -- 67, 1994.

\bibitem{Anantrasirichai:dynamic:2006}
N.~{Anantrasirichai}, C.~N. {Canagarajah}, D.~W. {Redmill}, and D.~R. {Bull},
\newblock ``Dynamic programming for multi-view disparity/depth estimation,''
\newblock in {\em IEEE International Conference on Acoustics Speech and Signal
  Processing Proceedings}, May 2006, vol.~2, pp. II--II.

\bibitem{Fan:Road:2018}
R.~{Fan}, X.~{Ai}, and N.~{Dahnoun},
\newblock ``Road surface 3d reconstruction based on dense subpixel disparity
  map estimation,''
\newblock {\em IEEE Transactions on Image Processing}, vol. 27, no. 6, pp.
  3025--3035, June 2018.

\bibitem{Ince:depth:2007}
S.~{Ince}, E.~{Martinian}, S.~{Yea}, and A.~{Vetro},
\newblock ``Depth estimation for view synthesis in multiview video coding,''
\newblock in {\em 2007 3DTV Conference}, May 2007, pp. 1--4.

\bibitem{Li:3DMIST:2017}
Lincheng Li, Xin Yu, Shunli Zhang, Xiaolin Zhao, and Li~Zhang,
\newblock ``3d cost aggregation with multiple minimum spanning trees for stereo
  matching,''
\newblock {\em Appl. Opt.}, vol. 56, pp. 3411--3420, 2017.

\bibitem{Scharstein:taxonomy:2002}
D.~Scharstein and R.~Szeliski,
\newblock ``A taxonomy and evaluation of dense two-frame stereo correspondence
  algorithms,''
\newblock {\em International Journal of Computer Vision}, vol. 47, pp. 7--42,
  2002.

\bibitem{Menze:Object:2015}
Moritz Menze and Andreas Geiger,
\newblock ``Object scene flow for autonomous vehicles,''
\newblock in {\em Conference on Computer Vision and Pattern Recognition
  (CVPR)}, 2015.

\bibitem{Atienza:Fast:2018}
R.~{Atienza},
\newblock ``Fast disparity estimation using dense networks,''
\newblock in {\em 2018 IEEE International Conference on Robotics and Automation
  (ICRA)}, May 2018, pp. 3207--3212.

\bibitem{Zbontar:Stereo:2016}
Jure \u{Z}bontar and Yann LeCun,
\newblock ``Stereo matching by training a convolutional neural network to
  compare image patches,''
\newblock {\em The Journal of Machine Learning Research}, vol. 17, no. 1, pp.
  2287--2318, 2016.

\bibitem{Chang:pyramid:2018}
J.~{Chang} and Y.~{Chen},
\newblock ``Pyramid stereo matching network,''
\newblock in {\em 2018 IEEE/CVF Conference on Computer Vision and Pattern
  Recognition}, June 2018, pp. 5410--5418.

\bibitem{Newell:hourglass:2016}
Alejandro Newell, Kaiyu Yang, and Jia Deng,
\newblock ``Stacked hourglass networks for human pose estimation,''
\newblock in {\em Computer Vision - ECCV 2016}, Bastian Leibe, Jiri Matas, Nicu
  Sebe, and Max Welling, Eds., Cham, 2016, pp. 483--499, Springer International
  Publishing.

\bibitem{Lu:sparse:2018}
Chuanhua Lu, Hideaki Uchiyama, Diego Thomas, Atsushi Shimada, and Rin ichiro
  Taniguchi,
\newblock ``Sparse cost volume for efficient stereo matching,''
\newblock {\em Remote sensing}, vol. 10, no. 11, pp. 1--12, 2018.

\bibitem{Zhang:GANet:2019}
Feihu Zhang, Victor Prisacariu, Ruigang Yang, and Philip~H.S. Torr,
\newblock ``{GA}-{N}et: {G}uided aggregation net for end-to-end stereo
  matching,''
\newblock {\em arXiv:1904.06587v1}, 2019.

\bibitem{Yu:Multi:2016}
F.~Yu and V.~Koltun,
\newblock ``Multi-scale context aggregation by dilated convolutions,''
\newblock in {\em International Conference on Learning Representations}, 2016.

\bibitem{Huang:Densely:2017}
G.~{Huang}, Z.~{Liu}, L.~v.~d. {Maaten}, and K.~Q. {Weinberger},
\newblock ``Densely connected convolutional networks,''
\newblock in {\em 2017 IEEE Conference on Computer Vision and Pattern
  Recognition (CVPR)}, July 2017, pp. 2261--2269.

\bibitem{Johnson:Perceptual:2016}
Justin Johnson, Alexandre Alahi, and Li~Fei-Fei,
\newblock ``Perceptual losses for real-time style transfer and
  super-resolution,''
\newblock in {\em European Conference on Computer Vision}, 2016.

\bibitem{Unity:2019}
Unity,
\newblock ``Unity real-time development platform,''
\newblock 2019.

\bibitem{Whelan:ElasticFusion:2016}
Thomas Whelan, Renato~F Salas-Moreno, Ben Glocker, Andrew~J Davison, and Stefan
  Leutenegger,
\newblock ``Elasticfusion: {R}eal-time dense slam and light source
  estimation,''
\newblock {\em The International Journal of Robotics Research}, vol. 35, no.
  14, pp. 1697--1716, 2016.

\bibitem{Butler:Naturalistic:2012}
Daniel~J. Butler, Jonas Wulff, Garrett~B. Stanley, and Michael~J. Black,
\newblock ``A naturalistic open source movie for optical flow evaluation,''
\newblock in {\em Computer Vision -- ECCV 2012}, Andrew Fitzgibbon, Svetlana
  Lazebnik, Pietro Perona, Yoichi Sato, and Cordelia Schmid, Eds., Berlin,
  Heidelberg, 2012, pp. 611--625, Springer Berlin Heidelberg.

\bibitem{Mayer:large:2016}
N.~Mayer, E.~Ilg, P.~H{\"a}usser, P.~Fischer, D.~Cremers, A.~Dosovitskiy, and
  T.~Brox,
\newblock ``A large dataset to train convolutional networks for disparity,
  optical flow, and scene flow estimation,''
\newblock in {\em IEEE International Conference on Computer Vision and Pattern
  Recognition (CVPR)}, 2016,
\newblock arXiv:1512.02134.

\end{thebibliography}

\end{document}